\documentclass[12pt]{article}
\usepackage{fancyhdr}

\usepackage{mathrsfs}
\usepackage[T1]{fontenc}
\usepackage{mathpazo}
\usepackage{setspace}
\usepackage{amsfonts}
\usepackage{amssymb}
\usepackage{amsmath}
\usepackage{epsfig}
\usepackage{latexsym}
\usepackage{color}
\usepackage{graphicx}
\usepackage{nicefrac}
\usepackage[latin1]{inputenc}
\usepackage{slashed}
\usepackage{multirow}
\usepackage{cite}
\usepackage{hyperref}
\usepackage{comment}

\def\hybrid{\topmargin -20pt    \oddsidemargin 0pt
        \headheight 0pt \headsep 0pt
        \textwidth 6.25in       
        \textheight 9.25in       
        \marginparwidth .875in
        \parskip 5pt plus 1pt   \jot = 1.5ex}

\hybrid

\def\baselinestretch{1.2}

\catcode`\@=11

\def\marginnote#1{}
\newcount\hour
\newcount\minute
\newtoks\amorpm
\hour=\time\divide\hour by60
\minute=\time{\multiply\hour by60 \global\advance\minute by-\hour}
\edef\standardtime{{\ifnum\hour<12 \global\amorpm={am}%
        \else\global\amorpm={pm}\advance\hour by-12 \fi
        \ifnum\hour=0 \hour=12 \fi
        \number\hour:\ifnum\minute<10 0\fi\number\minute\the\amorpm}}
\edef\militarytime{\number\hour:\ifnum\minute<10 0\fi\number\minute}

\def\draftlabel#1{{\@bsphack\if@filesw {\let\thepage\relax
   \xdef\@gtempa{\write\@auxout{\string
      \newlabel{#1}{{\@currentlabel}{\thepage}}}}}\@gtempa
   \if@nobreak \ifvmode\nobreak\fi\fi\fi\@esphack}
        \gdef\@eqnlabel{#1}}
\def\@eqnlabel{}
\def\@vacuum{}
\def\draftmarginnote#1{\marginpar{\raggedright\scriptsize\tt#1}}

\def\draft{\oddsidemargin -.5truein
        \def\@oddfoot{\sl preliminary draft \hfil
        \rm\thepage\hfil\sl\today\quad\militarytime}
        \let\@evenfoot\@oddfoot \overfullrule 3pt
        \let\label=\draftlabel
        \let\marginnote=\draftmarginnote
   \def\@eqnnum{(\theequation)\rlap{\kern\marginparsep\tt\@eqnlabel}%
\global\let\@eqnlabel\@vacuum}  }

\def\preprint{\twocolumn\sloppy\flushbottom\parindent 2em
        \leftmargini 2em\leftmarginv .5em\leftmarginvi .5em
        \oddsidemargin -.5in    \evensidemargin -.5in
        \columnsep .4in \footheight 0pt
        \textwidth 10.in        \topmargin  -.4in
        \headheight 12pt \topskip .4in
        \textheight 6.9in \footskip 0pt
        \def\@oddhead{\thepage\hfil\addtocounter{page}{1}\thepage}
        \let\@evenhead\@oddhead \def\@oddfoot{} \def\@evenfoot{} }

\def\numberbysection{\@addtoreset{equation}{section}
        \def\theequation{\thesection.\arabic{equation}}}

\def\underline#1{\relax\ifmmode\@@underline#1\else
        $\@@underline{\hbox{#1}}$\relax\fi}

\def\titlepage{\@restonecolfalse\if@twocolumn\@restonecoltrue\onecolumn
     \else \newpage \fi \thispagestyle{empty}\c@page\z@
        \def\thefootnote{\fnsymbol{footnote}} }

\def\endtitlepage{\if@restonecol\twocolumn \else \newpage \fi
        \def\thefootnote{\arabic{footnote}}
        \setcounter{footnote}{0}}  

\catcode`@=12
\relax

\def\figcap{\section*{Figure Captions\markboth
        {FIGURECAPTIONS}{FIGURECAPTIONS}}\list
        {Figure \arabic{enumi}:\hfill}{\settowidth\labelwidth{Figure
999:}
        \leftmargin\labelwidth
        \advance\leftmargin\labelsep\usecounter{enumi}}}
 \relax
\def\tablecap{\section*{Table Captions\markboth
        {TABLECAPTIONS}{TABLECAPTIONS}}\list
        {Table \arabic{enumi}:\hfill}{\settowidth\labelwidth{Table
999:}
        \leftmargin\labelwidth
        \advance\leftmargin\labelsep\usecounter{enumi}}}
 \relax
\def\reflist{\section*{References\markboth
        {REFLIST}{REFLIST}}\list
        {[\arabic{enumi}]\hfill}{\settowidth\labelwidth{[999]}
        \leftmargin\labelwidth
        \advance\leftmargin\labelsep\usecounter{enumi}}}
 \relax
\makeatletter
\newcounter{pubctr}
\def\publist{\@ifnextchar[{\@publist}{\@@publist}}
\def\@publist[#1]{\list
        {[\arabic{pubctr}]\hfill}{\settowidth\labelwidth{[999]}
        \leftmargin\labelwidth
        \advance\leftmargin\labelsep
        \@nmbrlisttrue\def\@listctr{pubctr}
        \setcounter{pubctr}{#1}\addtocounter{pubctr}{-1}}}
\def\@@publist{\list
        {[\arabic{pubctr}]\hfill}{\settowidth\labelwidth{[999]}
        \leftmargin\labelwidth
        \advance\leftmargin\labelsep
        \@nmbrlisttrue\def\@listctr{pubctr}}}
 \relax
\makeatother
\newskip\humongous \humongous=0pt plus 1000pt minus 1000pt

\newif\ifdtup

\relax

\def\be{\begin{equation}}
\def\ee{\end{equation}}
\def\ba{\begin{eqnarray}}
\def\ea{\end{eqnarray}}

\def\del{\partial}

\definecolor{markcolor2}{rgb}{1,0,0}

\def\b{\beta}

\def\d{\delta}
\def\D{\Delta}

\def\m{\mu}
\def\n{\nu}
\def\om{\omega}

\def\l{\lambda}
\def\L{\Lambda}
\def\s{\sigma}

\def\cN{{\cal N}}

\def\no{\noindent}

\def\qq{\qquad}

\def\IR{\relax{\rm I\kern-.18em R}}

\def \ha {{1\over 2}}

\def \ov {\over}

\def\IR{\relax{\rm I\kern-.18em R}}
\def\IL{\relax{\rm I\kern-.18em L}}

\def\inv{^{\raise.15ex\hbox{${\scriptscriptstyle -}$}\kern-.05em 1}}

\begin{document}

\renewcommand{\theequation}{\arabic{equation}}

\newcommand{\beq}{\begin{equation}}
\newcommand{\eeq}[1]{\label{#1}\end{equation}}
\newcommand{\ber}{\begin{eqnarray}}
\newcommand{\eer}[1]{\label{#1}\end{eqnarray}}
\newcommand{\eqn}[1]{(\ref{#1})}
\begin{titlepage}
\begin{center}


~
\vskip  .7in

{\large \bf The all-loop non-Abelian Thirring model and its RG flow}

\vskip 0.35in

{\bf Georgios Itsios},$^{1}$\phantom{x}{\bf Konstadinos Sfetsos}$^{2}$\ and\ {\bf Konstadinos Siampos}$^{3}$
\vskip 0.1in
{\em
\vskip .15in
${}^1$Department of Mathematics, University of Patras,
26110 Patras, Greece\\
{\tt gitsios@upatras.gr}\\
\vskip 0.1in
${}^2$Department of Physics, University of Athens,
15771 Athens, Greece\\
{\tt ksfetsos@phys.uoa.gr}\\
\vskip 0.1in
${}^3$M\'ecanique et Gravitation, Universit\'e de Mons, 7000 Mons, Belgique\\
{\tt konstantinos.siampos@umons.ac.be}
}

\vskip .5in
\end{center}

\centerline{\bf Synopsis}

\no
We analyze the renormalization group flow in a recently constructed class of integrable $\sigma$-models
which interpolate between WZW current algebra models and the non-Abelian T-duals of PCM for a simple group $G$.
They are characterized by the integer level $k$ of the current algebra, a deformation parameter $\l$ and they exhibit
a remarkable invariance involving the inversion of $\l$.
We compute the $\b$-function for $\l$ to leading order in $\displaystyle 1\ov k$.
Based on agreement with previous results for the exact $\b$-function of
the non-Abelian { bosonized} Thirring model and matching global symmetries,
we state that our integrable models are the resummed version (capturing all
counterterms in perturbation theory) of the non-Abelian { bosonized} Thirring model for a simple group $G$.
Finally, we present an analogous treatment in a simple example of a closely related class of models
interpolating between gauged WZW coset CFTs and the non-Abelian T-duals of PCM for the coset $G/H$.

\newpage


\end{titlepage}

\def\baselinestretch{1.2}
\baselineskip 20 pt
\noindent

\tableofcontents
\vskip .4in


\setcounter{equation}{0}
\renewcommand{\theequation}{\thesection.\arabic{equation}}
\section{Introduction and set-up}
\renewcommand{\theequation}{\thesection.\arabic{equation}}

A class of integrable $\s$-models with a group theoretical structure was recently
constructed explicitly in \cite{Sfetsos:2013wia} (using the algebraic construction set-up
in \cite{Balog:1993es}), which we first review.
Consider a general compact simple group $G$.
For a group element $g\in G$ parametrized by $X^\m$, $\m=1,2,\dots , \dim(G)$,
we introduce the right and left invariant Maurer--Cartan forms as follows
\be
J^a_+ = -i\, {\rm Tr}(t^a \del_+ g g^{-1}) = R^a_\m \del_+ X^\m \ ,\qq J^a_- = -i\, {\rm Tr}(t^a g^{-1} \del_- g )= L^a_\m \del_- X^\m\ ,
\ee
where the matrices $t^a$ obey the commutation relations $[t_a,t_b]=i f_{abc} t_c$
and are normalized as ${\rm Tr}(t_a t_b)=\d_{ab}.$ Hence, there is no difference between upper and lower tangent space indices.
The Maurer--Cartan forms are related by an orthogonal matrix $D$ as
\be
R^a = D_{ab}L^b \ ,\qq D_{ab}={\rm Tr}(t_a g t_b g^{-1})\ .
\ee
Then the form of the integrable $\s$-model action is\footnote{
The action bellow is the simplest one of a class of multi-parameter models constructed in \cite{Sfetsos:2013wia}.
However, only in special cases, such as the one below, these $\s$-models are expected to be integrable.
}
\be
S_{k,\l}(g) = S_{{\rm WZW},k}(g) + {{k \l}\ov \pi}  \int J_+^a (\mathbb{1}-\l D^T)^{-1}_{ab}J_-^b  \ ,
\label{tdulalmorev2}
\ee
where
\be
S_{{\rm WZW},k}(g) = -{k\ov 2\pi} \int {\rm Tr}(g^{-1} \del_+ g g^{-1} \del_- g) + {i k\ov 6\pi} \int_B {\rm Tr}(g^{-1}\mathrm{d}g)^3\ ,
\label{dhwzw}
\ee
is the Wess--Zumino--Witten (WZW) action at level $k$  and $\l$ is a
real coupling constant.

\no
These models were constructed through a gauging procedure and are invariant under the global symmetry
$g\to \L_0^{-1} g\L_0$, where $\L_0\in G$. Moreover, their coupling constant is
$0\leqslant \l \leqslant 1$ by construction.
For $\l\ll 1$ the action \eqn{tdulalmorev2} corresponds to the WZW theory perturbed by the current bilinear
term as
\be
S_{k,\l}(g) = S_{{\rm WZW},k}(g) + {k\l\ov \pi} \int J_+^a J_-^a + {\cal O}(\l^2)\ ,
\label{thirr}
\ee
which clearly preserves the above global symmetry. The first two terms in the above expansion define the
so-called non-Abelian {bosonized} Thirring model ({in short non-Abelian Thirring model}) \cite{Dashen:1974gu}{,
see also \cite{Karabali:1988sz}}. For this model the $\b$-function for $\l$ has been computed,
to leading order in the $1/k$ expansion, but exactly in $\l$. The result is \cite{Kutasov:1989dt}
\be
{\mathrm{d}\l\ov \mathrm{d}t} = -\frac{c_G\l^2}{2k(1+\l)^2}\ ,
\label{betall}
\ee
where $t=2\pi \ln \m$, with $\m$ being the energy scale and
where $c_G$ is the quadratic Casimir in the adjoint representation, defined from the relation $f_{acd}f_{bcd}= c_G \d_{ab}$.
Note that this equation is invariant under the transformation $\l\to 1/\l$ and $k\to -k$, which is a symmetry of \eqn{tdulalmorev2} in a way that is made precise in \eqn{symmetry.sigma} below.
Moreover this map exists also in the non-Abelian Thirring model for large values of $k$ \cite{Kutasov:1989aw}.
In that sense \eqn{tdulalmorev2} captures all counterterms in perturbation
theory corresponding to the coupling $\l$ and to leading order in $1/k$.\footnote{It is tempting to
suggest that the exact to all orders in $1/k$ action is given by \eqn{tdulalmorev2}, but with $k$ replaced by $k+c_G$.
This replacement is in accordance with the exact map $k\to -k -2 c_G$ of \cite{Kutasov:1989aw} we mentioned above.}
We also note that the necessary conditions for one-loop conformal invariance of a general class of models, which includes 
\eqn{tdulalmorev2}, were derived in \cite{Tseytlin:1993hm}.

\no
For $\l \to 1$ the $\s$-model action is effectively described by the non-Abelian T-dual of the Principal Chiral Model (PCM)
for the group $G$.\footnote{For recent developments and the usage of non-Abelian T-duality in supergravity,
string theory and the gauge/gravity correspondence, as well as additional references in the literature, the reader is advised to
consult \cite{Sfetsos:2010uq,Lozano:2011kb,Itsios:2013wd}.}
The correspondence involves a limiting procedure for the coordinates $X^\m$ parametrizing the group element $g\in G$ and the details can be found in \cite{Sfetsos:2013wia}.
The value $\l=1$ is special since once crossed from below the $\s$-model metric changes its signature from Euclidean
by picking up an overall sign. It is also a self-dual point of the above $\l\to 1/\l$ transformation.

\section{Renormalization group flow restricted by symmetries}
\setcounter{equation}{0}
\renewcommand{\theequation}{\thesection.\arabic{equation}}

The overall coupling constant $k$ is not expected, being an integer, to get renormalized,
a fact that will be confirmed by our computation. In contrast, the coupling constant
$\l$ is expected to have a non-trivial running since the perturbation
$J_+^aJ_-^a$ is not exactly marginal. The purpose of this section
is to restrict the form of the corresponding $\b$-function
$\displaystyle \b_\l=\m {\mathrm{d}\l\ov \mathrm{d}\m}$ by symmetry considerations.
In the next section we will explicitly compute the $\b$-function and prove that it is compatible with
symmetry arguments.

\no
It is useful to extend the range of the coupling constant $\l$ so that $0\leqslant \l < \infty$.
Then the following remarkable property
\be
\label{symmetry.sigma}
S_{-k,1/\l}(g^{-1}) = S_{k,\l}(g)\ ,
\ee
holds true. This implies a large/small coupling duality under a simultaneous flipping
of the sign of the overall coupling $k$. This duality severely restricts the form of the RG flow
equation for $\l$. This equation is of the form
\be
\b_\l = \m{\mathrm{d}\l\ov \mathrm{d}\m} = - {1\ov 2\pi} {f(\l)\ov  k} \ ,
\ee
where $f(\l)$ is a function to be determined. Due to the above duality symmetry the relation
\be
f(1/\l) = \l^{-2} f(\l) \ ,
\ee
should hold, which severely constrains the function $f(\l)$.
From the structure of the action \eqn{tdulalmorev2} and in particular
the fact that it is built up by finite dimensional matrices,
it is clear that the function $f(\l)$ should be the ratio of two polynomials.
The coefficients of these polynomials can be almost completely determined as follows:
Let us first recall that when perturbing a CFT by terms of the form $\l_i \Phi_i$, where the operators $\Phi_i$ have anomalous dimensions $\D_i$,
the $\beta$-functions for the couplings $\l_i$, up to two-loops in perturbation theory, are of the form (see, for instance, \cite{bookJustin})
\be
{\mathrm{d}\l_i\ov \mathrm{d}t} = -(2-\D_i)\l_i - C_i{}^{jk} \l_j \l_k\ + {\cal O}(\l^3)\ ,
\label{dlt}
\ee
where $C_{ijk}$ are the coefficients of the operator product expansions of the operators $\Phi_i$ among themselves.
In our case we have a single operator, namely that $\Phi_1= J_+^a J_-^a$ with $\D_1=2$.
Using that the $J_\pm^a$'s obey two mutually commuting current algebras,
we easily compute that $C_i{}^{11}= c_G \d_{i,1}$, where, as noted, $c_G$ is the quadratic Casimir in the adjoint representation.
That means in our case $\displaystyle {\mathrm{d}\l \ov \mathrm{d}t} = - c_G \l^2 +   {\cal O}(\l^3)$.
Then the function $f(\l)$ will be the ratio of two polynomials
whose degrees as well as their coefficients, for each one of them separately, will be related
via to the above large/small coupling duality symmetry.
Clearly, if we know the structure of the zeros and the poles of $f(\l)$ we can determine (almost)
completely the RG flow equation for $\l$. We know that there is only one conformal point in which
$\b_\l=0$, i.e. when $\l\to 0$, reached in the UV.
Therefore $f(\l)$ cannot have any zeros for real $\l$.
There is also no reason to reach a conformal point for $\l$ complex. Hence, we end up with the expression
\be
f(\l) = -{c_G \l^2\ov 1+ a \l + \l^2} \ ,
\ee
for some constant $a$, which clearly exhibits the correct perturbative behaviour.
The form of the background's details as discussed in appendix A suggests that $a=\pm 2$.
Of course continuing with such type of arguments can leave unsatisfied a skeptical reader. In the next section we will take up the task of actually
explicitly proving \eqn{betall}, a result corresponding to the value of the constant $a=2$.

\section{Tour de force}
\setcounter{equation}{0}
\renewcommand{\theequation}{\thesection.\arabic{equation}}

In this section we explicitly compute the $\b$-function for $\l$ using the form of the background.

\no
Being an integer we expect that $k$ is {not} running with the energy scale. {
This is not an assumption as we are going to prove it at the end of this section.}
It is convenient to write the metric using a frame $e^a=e^a_\m \mathrm{d} X^\m$,
as $g_{\m\n}= e^a_\m e^a_\n$. Any other frame $\tilde e^a$ will be related to this one by an orthogonal
transformation, i.e. $\tilde e^a = \L^{ab} e^b$.
In our case these are given by \cite{Sfetsos:2013wia}\footnote{We reinstate the overall factor $\sqrt{k(1-\l^2)}$ as
compared to the correspondent expression in \cite{Sfetsos:2013wia}.}
\be
e^a=\sqrt{k(1-\l^2)}\,(D-\l\mathbb{I})^{-1}_{ab}R^b\ ,\qq  \L = {D-\l \mathbb{1}\ov \mathbb{1} - \l D}\ .
\label{ehfjfr}
\ee
The one-loop $\b$-function equations are given by \cite{honer,Friedan:1980jf,Curtright:1984dz}
\be
\label{RG.flows1}
{\mathrm{d} g_{\mu\nu}\ov \mathrm{d}t}-{\mathrm{d} B_{\mu\nu}\ov \mathrm{d}t}=R^+_{\mu\nu}+\nabla^+_\nu\xi_\mu\ .
\ee
where the second term corresponds to diffeomorphisms along $\xi^\m$.
Passing to the tangent space indices and using the definitions
\be
{\mathrm{d} g_{\mu\nu}\ov \mathrm{d}t} = \b^g_{ab} e^a_\m e^b_\n\ ,\qq  {\mathrm{d} B_{\mu\nu}\ov \mathrm{d}t} = \b^B_{ab} e^a_\m e^b_\n\ ,
\ee
we have that
\be
\b^g_{ab}-\b^B_{ab}=R^+_{ab}+\nabla^+_b\xi_a\ ,
\label{RG.flows}
\ee
We next compute the left hand side of this equation.
Since the WZW model first term in \eqn{tdulalmorev2} does not depend on the parameter $\l$ we immediately have that
\be
\begin{split}
&  { \mathrm{d} {g}_{\m\n}\ov  \mathrm{d}t} + { \mathrm{d} B_{\m\n}\ov  \mathrm{d}t}
=   2 k { \mathrm{d}\ov  \mathrm{d}t}\left(\l R^a_\m (1-\l D^T)^{-1}_{ab} L^b_\n\right)
\\
&\phantom{xxxxxxxxxxx}  =  2 {k} {d\l\ov d t} R^a_\m \left[(1-\l\ D^T)^{-1} (1-\l D^T)^{-1}\right]_{ab} L^b_\n\ ,
\\
&
\phantom{xxxxxxxxxxx}  =  {2\ov 1-\l^2} { \mathrm{d}\l\ov  \mathrm{d}t} e^a_\m \L_{ba} e^b_\n\ ,
\end{split}
\ee
where in the last step we used the definition of the frame \eqn{ehfjfr}.
Then by letting the group element $g\to g^{-1}$ which reverses the sign of $B_{\m\n}$, we obtain that
\be
{ \mathrm{d} {g}_{\m\n}\ov  \mathrm{d}t} - { \mathrm{d} B_{\m\n}\ov  \mathrm{d}t}
= {2\ov 1-\l^2} { \mathrm{d}\l\ov  \mathrm{d}t} \tilde e^a_\m \L_{ab} \tilde e^b_\n = {2\ov 1-\l^2} { \mathrm{d}\l\ov  \mathrm{d}t}  e^a_\m \L_{ab} e^b_\n \ ,
\ee
from which
\be
\label{der.dif}
\b^g_{ab}-\b^B_{ab}=\frac{2}{1-\l^2}{\mathrm{d}\l\ov \mathrm{d}t}\L_{ab}\ .
\ee
The right hand side of the one-loop RG flow equation \eqn{RG.flows} can also be worked out. Indeed,
by choosing for $\xi_a=-c_2\,f_{abc}\L_{bc}$ and using \eqref{dLambda}, \eqref{spinp} and \eqref{Ricci},
we can prove that
\be
\label{Ricci.diff}
R^+_{ab}+\nabla^+_b\xi_a = - c_G\,c_2^2\,\L_{ab}\ .
\ee
Plugging \eqref{Ricci.diff} and \eqref{der.dif} in \eqref{RG.flows} and using the expression \eqref{cs} for the constant $c_2$,
we readily find that the RG flow equation reads
\be
{\mathrm{d}\l\ov \mathrm{d}t} = -\frac{c_G\l^2}{2k(1+\l)^2}\ ,
\label{betalla}
\ee
which is nothing but \eqn{betall}.\footnote{It turns out that the system of beta-function equations
computed for the $SU(2)$ case in \cite{Balog:1993es} and \cite{Evans:1994hi} is consistent with \eqn{betalla}.}
This is a quite simple formula, valid for all simple compact groups and constitutes one of the main results of present paper.
It is essentially universal in the sense that its dependence on the group is
only through the overall coefficient $c_G$. In fact \eqn{betalla} can be solved explicitly, leading to
\be
\l-{1\ov \l} + \ln \l^2 = -{c_g\ov 2 k} (t-t_0)\ ,
\ee
where $t_0$ is an integration constant. In the UV at $t\to \infty$, we have that $\l\to 0$ and one reaches the
conformal point described by the WZW action. Towards the IR at $t=t_0$ one reaches the
self-dual point $\l=1$ corresponding to the non-Abelian T-dual of the PCM as mentioned above.

\no
As was discussed, the form of the RG flow equations is such that $k$ does not run. Its quantization of topological nature
\cite{Witten:1983ar} remains
unaltered at one-loop, a fact which is expected to hold true to all orders in perturbation theory. For completeness we note that
if we had not assumed that $k$ would remain fixed, we would have obtained instead of \eqn{der.dif} that
\be
\b^g_{ab}-\b^B_{ab}=\frac{2}{1-\l^2}{\mathrm{d}\l\ov \mathrm{d}t}\L_{ab}+{1\ov k} {\mathrm{d} k\ov \mathrm{d}t}(\d_{ab}-b_{ab})\ ,
\ee
where $b$ is a matrix defined from the antisymmetric two-form as $B_{\m\n} =b_{ab} e^a_\m e^b_\n$. Clearly,
only by requiring that $\displaystyle {\mathrm{d}k\ov \mathrm{d}t}=0$ we can match with \eqn{Ricci.diff}.

\section{Renormalization group flow on cosets}
\setcounter{equation}{0}
\renewcommand{\theequation}{\thesection.\arabic{equation}}

Closely related to \eqn{tdulalmorev2} there is a class of models interpolating between exact coset $G/H$ CFT
realized by gauged WZW models and the non-Abelian T-duals of the PCM for the geometric coset $G/H$ spaces \cite{Sfetsos:2013wia}.
These models have not been shown to be integrable, though we expect integrability for the cases that $G/H$ is a symmetric space.
For the case of $G=SU(2)$ and $H=U(1)$ the details have been worked out \cite{Sfetsos:2013wia}. The result
for the $\s$-model action can be presented as
\be
\begin{split}
& S
= {k\ov \pi} \int \Big[  {1-\l\ov 1+\l} \left(\del_+\om \del_-\om + \cot^2\om \del_+ \phi\del_- \phi\right)\\
& \phantom{}
\phantom{xxx} + 4 {\l\ov 1-\l^2}
(\cos\phi \del_+ \om + \sin\phi \cot\om \del_+ \phi)
(\cos\phi \del_- \om + \sin\phi \cot\om \del_- \phi)\Big]\ .
\label{clpp2}
\end{split}
\ee
This action is invariant under the large/small duality symmetry for which $\l\to 1/\l$ and $k\to -k$.
It has been shown in \cite{Sfetsos:2013wia} that for $\l\ll 1$ this represents the corresponding $\s$-model action
for the coset $SU(2)_k/U(1)$ CFT perturbed by the parafermion bilinears
${k\l\ov \pi} \int (\psi\bar \psi + \psi^\dagger \bar \psi^\dagger)$.

\no
In two target space dimensions the one-loop RG flow equation is simply given by
\be
{\mathrm{d}g_{\m\n}\ov \mathrm{d}t} = {R\ov 2} g_{\m\n} + \nabla_\m \xi_\n +  \nabla_\n \xi_\m\ .
\ee
It turns that the above $\s$-model is one-loop renormalizable and the corresponding RG flow equation for $\l$ is simply given by
\be
{\mathrm{d}\l\ov \mathrm{d}t} = -\frac{2\l}{k}\ ,
\ee
where we also found necessary to employ a diffeomorphism with $\xi_\omega= - \cot\omega $ and $\xi_{\phi} =0$.
It is remarkable that this result coincides to the one-loop perturbative result in $\l$. This follows directly from the
general expression \eqref{dlt} with scaling dimension $2-2/k$ and, as it turns out, vanishing operator product expansion structure constants.
It is also invariant under the large/small symmetry $\l\to 1/\l$ and $k\to -k$, as expected. In the UV one reaches the exact $SU(2)_k/U(1)$
CFT and towards the IR at $\l=1$ the theory corresponds to the non-Abelian T-dual of $S^2$ with respect to $SU(2)$ via a limiting procedure
involving also the coordinates $\om$ and $\phi$.
The details can be found in \cite{Sfetsos:2013wia}.

\section{Concluding remarks and outlook}
\setcounter{equation}{0}

We have computed the one-loop renormalization group flow for the integrable $\sigma$-model action
\eqn{tdulalmorev2} interpolating between WZW current algebra models and the non-Abelian T-duals of PCM for a group $G$.
The $\b$-function for the deformation parameter $\l$ coincided with that computed in the past, and argued to be exact,
for the non-Abelian Thirring model. Based on the fact that the two models have the same global symmetries it is
natural to suggest that the $\s$-model action \eqn{tdulalmorev2} is a resummed version of the non-Abelian Thirring model
action (given by the first two terms in \eqn{thirr}) in which all perturbative, in the deformation parameter $\l$, effects
have been taken into account.
To further support our suggestion one could compute using the general results of \cite{Sfetsos:2013wia} the analog
of the $\s$-model action \eqn{tdulalmorev2} but with more than one deformation parameters such that
when they are small it yields the form of an ``anisotropic" non-Abelian Thirring model
\begin{equation}
S_{k,\l}(g) = S_{{\rm WZW},k}(g) + {k\ov \pi} \sum_{a=1}^{\dim G}\l_a\int J_+^a J_-^a + {\cal O}(\l_a^2)\ .
\end{equation}
The result for the running of the $\l_a$'s under the renormalization group flow can then be
compared to that obtained in \cite{Gerganov:2000mt} for the case of the $SU(2)$ group for
the "anisotropic" non-Abelian Thirring model with couplings $\l_1=\l_2\neq \l_3$.

\no
Finally, we note that the integrable models described by the $\s$-model action \eqn{tdulalmorev2} are
distinct from the ones constructed in \cite{Klimcik:2008eq}, for which the running of the deformation
parameter was computed in \cite{Squellari:2014jfa}.

\section*{Acknowledgements}

We would like to thank A. Torrielli for a useful correspondence. The
research of G. Itsios has been co-financed by the ESF (2007-2013) and Greek
national funds through the Operational Program ``Education and
Lifelong Learning" of the NSRF - Research Funding Program:
``Heracleitus II. Investing in knowledge in society through the
European Social Fund". The research of K.\,Sfetsos is implemented
under the \textsl{ARISTEIA} action (D.654 GGET) of the \textsl{operational
programme education and lifelong learning} and is co-funded by the
European Social Fund (ESF) and National Resources (2007-2013). The work of K.
Siampos has been supported by  \textsl{Actions de recherche
concert\'ees (ARC)} de la \textsl{Direction g\'en\'erale de
l'Enseignement non obligatoire et de la Recherche scientifique
Direction de la Recherche scientifique Communaut\'e fran\c{c}aise de
Belgique} (AUWB-2010-10/15-UMONS-1), and by IISN-Belgium (convention 4.4511.06). The authors
would like to thank each others home institutions for hospitality
and financial support, where part of this work was developed.

\appendix
\section{The generalized curvature and Ricci tensors}

\setcounter{equation}{0}
\renewcommand{\theequation}{\thesection.\arabic{equation}}

In this appendix we derive the expressions for the generalized Riemann and Ricci tensors
constructed using the torsion. We follow the conventions of \cite{Sfetsos:2013wia}.

\no
The torsionless and metric compatible spin connection is defined by
\be
\mathrm{d} e^a + \om_{ab} \wedge e^b = 0\ ,\qq \om_{ab}=-\om_{ba}\,.
\ee
The torsionless Riemann 2-forms are constructed as
\be
\Omega_{ab} = \mathrm{d}\om_{ab} +  \om_{ac} \wedge \om_{cb} = \ha R_{ab|cd}\,e^c\wedge e^d\ ,
\ee
In addition, the definition of the torsionfull Riemann 2-forms is
\be
\Omega^\pm_{ab} = \mathrm{d}\om^\pm_{ab} + \om^\pm_{ac} \wedge \om^\pm_{cb} = \ha R^\pm_{ab|cd}\,e^c\wedge e^d\ .
\ee
We also use the symbols $\om_{ab|c}$ and $\om^\pm_{ab|c}$ defined by
\be
\begin{split}
& \om_{ab} = \om_{ab|c} \, e^c \ ,\qq \om^\pm_{ab} = \om^\pm_{ab|c} \, e^c\\
& \om^\pm_{ab|c} = \om_{ab|c} \pm \frac{1}{2} H_{abc}\ .
\end{split}
\ee

\no
Using the above conventions we can rewrite $\Omega^+_{ab}$ in the following form
\be
\Omega^+_{ab} = \mathrm{d}\om^+_{ab|d} \wedge e^d \, + \, \om^+_{ab|f} \Big( \om^+_{fe|d} - \frac{1}{2} H_{fed} \Big) \, e^e \wedge e^d \,
+ \, \om^+_{ac|e} \, \om^+_{cb|d} \, e^e \wedge e^d \ .
\ee
In our case the components $\om^{+}_{ab|c}$ are given by
\be
\label{spinp}
\om^{+}_{ab|c} = -c_2 \, f_{abd} \, \L_{dc} \ ,
\ee
while the components of $H={1\ov 6} H_{abc}\, e^a \wedge e^b \wedge e^c$ are
\be
\label{Hform}
H_{abc} = - c_1 \, f_{abc} - c_2 \, \Big( \L_{da} \, f_{dbc} +  \L_{db} \, f_{dca} +  \L_{dc} \, f_{dab} \Big) \ .
\ee
In order to compute $\Omega^+_{ab}$ we also need the expression
\be
\label{dLambda}
\mathrm{d}\L_{ab} = c_1 \, \L_{ae} \, f_{ebc} \, e^c + c_2 \, \Big(f_{abc} - f_{adc} \, \L_{db} + \L_{ae} \, f_{edc} \, \L_{db} \Big) \, e^c \, .
\ee
where
\be
\label{cs}
c_1 = \frac{1}{\sqrt{k(1-\l^2)}}\,{1 + \l + \l^2 \ov 1 + \l} \ ,\qq c_2 =\frac{1}{\sqrt{k(1-\l^2)}}\, {\l \ov 1 + \l} \, .
\ee
After some algebra the torsionfull Riemann 2-form is found to be
%
%
\be
\label{Omega2form}
\begin{split}
&\Omega_{ab}^+=\frac12\left(c_2^2\,f_{abe}f_{ecd}+c_1c_2\,f_{abe}f_{cdf}\L_{ef}+2c_2^2\,f_{abe}f_{efc}\L_{fd}\right.\\
&\left.-2c_2^2\,f_{abe}f_{fgc}\L_{ef}\L_{gd}-c_2^2\,f_{abe}f_{efg}\L_{fc}\L_{gd}\right)\,e^c\wedge e^d\,.
\end{split}
\ee
From the latter we can read off the generalized Riemann tensor
%
%
\be
\begin{split}
R^+_{ab|cd} &= c_2^2 \, f_{abe} \, f_{ecd}+c_1 \, c_2 \, f_{abe} \, f_{cdf} \, \L_{ef}
                 + c_2^2 \,f_{abe} \Big(f_{efc}\,\L_{fd} -f_{efd}\,\L_{fc} \Big) \, \\
                 & - c_2^2 \,f_{abe} \Big(\left(f_{fgc} \, \L_{gd}\,   \,  - f_{fgd} \, \L_{gc}\,\right) \L_{ef} \,
                 + \, f_{efg}\L_{fc} \, \L_{gd} \, \Big)\,.
\end{split}
\ee
Then one computes the generalized Ricci tensor. The result is given by
\be
\begin{split}
\label{Ricci}
R^+_{ab} &= c_G \, c_2^2 \, \Big( \d_{ab} - \L_{ab} \Big) + f_{age} \, f_{bce} \, \Big(c_1 \, c_2 \, \L_{gc} - c_2^2 \, \L_{cg}\Big)
 \\
           &\phantom{xx} + c_2^2 \, \Big( f_{ahe}\, f_{bcg}\,\L_{hc} \, \L_{ge} \, + f_{age}\, f_{hce}\,\L_{gb} \, \L_{hc} \,  \Big)\,.
\end{split}
\ee
Equivalently the latter could be computed directly through the torsionfull Riemann 2-forms \eqref{Omega2form}, i.e. $e^c\lrcorner\,\Omega^+_{ca}=R^+_{ab}\,e^b$.



\begin{thebibliography}{1}

 \bibitem{Sfetsos:2013wia}
  K.~Sfetsos, {\it Integrable interpolations: From exact CFTs to non-Abelian T-duals},\hfill\break
  Nucl. Phys. {\bf B880} (2014) 225, \href{http://arxiv.org/abs/arXiv:1312.4560}{arXiv:1312.4560 [hep-th]}.

\bibitem{Balog:1993es}
  J.~Balog, P.~Forgacs, Z.~Horvath and L.~Palla,
  {\it A New family of SU(2) symmetric integrable sigma models},
  Phys. Lett. {\bf B324} (1994) 403, \href{http://arxiv.org/abs/hep-th/9307030}{hep-th/9307030}.


\bibitem{Dashen:1974gu}
  R.F.~Dashen and Y.~Frishman,
  {\it Thirring model with u(n) symmetry - scale invariant only for fixed values of a coupling constant},
  \href{http://www.sciencedirect.com/science/article/pii/0370269373901615}{Phys.\ Lett. {\bf B46} (1973) 439}
  and
  {\it Four Fermion Interactions and Scale Invariance},
  \href{http://journals.aps.org/prd/abstract/10.1103/PhysRevD.11.2781}{Phys. Rev. {\bf D11} (1975) 2781.}
  
{ 

\bibitem{Karabali:1988sz}
  D.~Karabali, Q. H.~Park and H.~J.~Schnitzer,
  {\it Thirring Interactions, Nonabelian Bose-fermi Equivalences and Conformal Invariance},
 \href{http://www.sciencedirect.com/science/article/pii/0550321389901247}{Nucl.\ Phys.\ B {\bf 323} (1989) 572.}
}

\bibitem{Kutasov:1989dt}
  D.~Kutasov,
  {\it String Theory and the Nonabelian Thirring Model},\hfill\break
 \href{http://www.sciencedirect.com/science/article/pii/0370269389912859}{Phys. Lett. {\bf B227} (1989) 68}.

\bibitem{Kutasov:1989aw}
  D.~Kutasov, {\it Duality Off the Critical Point in Two-dimensional Systems With Nonabelian Symmetries},
  \href{http://www.sciencedirect.com/science/article/pii/0370269389913257}{Phys. Lett. {\bf B233} (1989) 369.}

\bibitem{Tseytlin:1993hm}
  A.A.~Tseytlin,
  {\it On a 'Universal' class of WZW type conformal models},
  Nucl. Phys. {\bf B418} (1994) 173,
 \href{http://arxiv.org/abs/hep-th/9311062}{hep-th/9311062}.


\bibitem{Sfetsos:2010uq}
  K.~Sfetsos and D.C.~Thompson,
  {\it On non-abelian T-dual geometries with Ramond fluxes},
  Nucl. Phys. {\bf B846} (2011) 21, \href{http://arxiv.org/abs/arXiv:1012.1320}{arXiv:1012.1320 [hep-th]}.

\bibitem{Lozano:2011kb}
  Y.~Lozano, E. O'Colgain, K.~Sfetsos and D.C.~Thompson,
  {\it Non-abelian T-duality, Ramond Fields and Coset Geometries},\hfill\break
  JHEP {\bf 1106} (2011) 106, \href{http://arxiv.org/abs/arXiv:1104.5196}{arXiv:1104.5196 [hep-th]}.

\bibitem{Itsios:2013wd}
  G.~Itsios, C.~Nunez, K.~Sfetsos and D.C.~Thompson,
  {\it Non-Abelian T-duality and the AdS/CFT correspondence:new $\cN=1$ backgrounds},\hfill\break
  Nucl. Phys. {\bf B873} (2013) 1, \href{http://arxiv.org/abs/arXiv:1301.6755}{arXiv:1301.6755 [hep-th]}.

\bibitem{Klimcik:2008eq}
  C. Klimcik,
{\it On integrability of the Yang-Baxter sigma-model},\hfill\break
  J. Math. Phys. {\bf 50} (2009) 043508,
\href{http://arxiv.org/abs/arXiv:0802.3518}{arXiv:0802.3518 [hep-th]}.


\bibitem{Squellari:2014jfa}
  R.~Squellari,
  {\it Yang-Baxter $\sigma$ model: Quantum aspects}
  Nucl. Phys. {\bf B881} (2014) 502, \href{http://arxiv.org/abs/arXiv:0802.3518}{arXiv:1401.3197 [hep-th]}.

\bibitem{bookJustin}
J. Cardy, in: E.~Brezin, J. Zinn-Justin (Eds.), Fields, {\it Strings and
Critical Phenomena}, Les Houches Lectures, 1989.

\bibitem{Evans:1994hi}
  J.M.~Evans and T.~J.~Hollowood,
  {\it Integrable theories that are asymptotically CFT},\hfill\break
  Nucl. Phys. {\bf B438} (1995) 469, \href{http://arxiv.org/abs/hep-th/9407113}{hep-th/9407113}.

\bibitem{Witten:1983ar}
  E.~Witten,
  {\it Nonabelian Bosonization in Two-Dimensions},\hfill\break
\href{http://link.springer.com/article/10.1007\%2FBF01215276}{Commun. Math. Phys. {\bf 92} (1984) 455}.


  \bibitem{honer}
 G.~Ecker and J.~Honerkamp,
 {\it Application of invariant renormalization to the nonlinear chiral invariant
 pion Lagrangian in the one-loop approximation},\hfill\break
 \href{http://www.sciencedirect.com/science/article/pii/0550321371904688}{Nucl. Phys. {\bf B35}, 481 (1971).}\hfill\break
J.~Honerkamp,
 {\it Chiral multiloops},
\href{http://www.sciencedirect.com/science/article/pii/0550321372902994}{Nucl. Phys. {\bf B36} (1972) 130.}


\bibitem{Friedan:1980jf}
  D.~Friedan,
  {\it Nonlinear Models in Two Epsilon Dimensions},\hfill\break
  \href{http://journals.aps.org/prl/abstract/10.1103/PhysRevLett.45.1057}{Phys. Rev. Lett. {\bf 45} (1980) 1057}
 and {\it Nonlinear Models in Two + Epsilon Dimensions},
  \href{http://www.sciencedirect.com/science/article/pii/0003491685903847}{Annals Phys.\  {\bf 163} (1985) 318.}
  
  \bibitem{Curtright:1984dz}
  T.~L.~Curtright and C.~K.~Zachos,
  {\it Geometry, Topology and Supersymmetry in Nonlinear Models},
\href{http://journals.aps.org/prl/abstract/10.1103/PhysRevLett.53.1799}{Phys.\ Rev.\ Lett.\  {\bf 53} (1984) 1799.}\hfill\break
 {
  E.~Braaten, T.~L.~Curtright and C.~K.~Zachos,
  {\it Torsion and Geometrostasis in Nonlinear Sigma Models},
  \href{http://www.sciencedirect.com/science/article/pii/0550321385900537}{Nucl.\ Phys.\ B {\bf 260} (1985) 630.}\hfill\break
  }
  B.E.~Fridling and A.E.M.van de Ven,
  {\it Renormalization of Generalized Two-dimensional Nonlinear $\sigma$-Models},
\href{http://www.sciencedirect.com/science/article/pii/0550321386902671}
{Nucl. Phys. {\bf B268} (1986) 719}.


\bibitem{Gerganov:2000mt}
  B.~Gerganov, A.~LeClair and M.~Moriconi,
  {\it On the beta function for anisotropic current interactions in 2-D},
  Phys. Rev. Lett. {\bf 86} (2001) 4753,
 \href{http://arxiv.org/abs/hep-th/0011189}{hep-th/0011189}.




\end{thebibliography}
\end{document}